\documentclass[12pt]{article}

\usepackage{amsmath,amssymb}
\usepackage{graphicx,a4,psfrag,here,epsfig,epsf}
\usepackage{pstricks}
\usepackage{pst-node}
\usepackage{citesort}
\usepackage{multirow}

\allowdisplaybreaks[1]

\begin{document}
\newcommand{\df}{\delta_1}
\newcommand{\ds}{\delta_2}
\newcommand{\se}{s_\varepsilon}
\newcommand{\ce}{c_\varepsilon}
\newcommand{\tf}{\tau_1}
\newcommand{\ts}{\tau_2}
\newcommand{\eq}{\begin{eqnarray}}
\newcommand{\en}{\end{eqnarray}}

\begin{titlepage}

\vspace{1cm}
\begin{center}{\Large\bf The {\boldmath$B\to K^*$} form factors on the lattice}

\vspace{0.5cm}
\today

\vspace{0.5cm}
Andria Agadjanov$^{a}$, V\'eronique Bernard$^b$, Ulf-G. Mei{\ss}ner$^{a,c}$
and Akaki Rusetsky$^a$

\vspace{2em}
\footnotesize{\begin{tabular}{c}
$^a\,$ Helmholtz-Institut f\"ur Strahlen- und Kernphysik (Theorie) and\\
Bethe Center for Theoretical Physics,\\
 $\hspace{2mm}$   Universit\"at Bonn, D-53115 Bonn, Germany\\[2mm]
$^b\,$Institut de Physique Nucl\'eaire, UMR 8608, CNRS, Univ. Paris-Sud,\\
           Universit\'e Paris-Saclay, F-91406 Orsay Cedex, France\\[2mm]
$^c\,$ 
Institute for Advanced Simulation (IAS-4), Institut f\"ur Kernphysik 
(IKP-3) and\\ J\"ulich Center for Hadron Physics,
Forschungszentrum J\"ulich, D-52425 J\"ulich, Germany
\end{tabular}  }

\vspace{1cm}

\begin{abstract}
\noindent
The extraction of the $B\to K^*$ transition form factors from lattice data is studied, applying non-relativistic effective field theory in a finite volume. The possible mixing of $\pi K$ and $\eta K$ states is taken into account. The two-channel analogue of the Lellouch-L\"uscher formula is reproduced. Due to the resonance nature of the $K^*$, an equation is derived, which allows to determine the form factors at the pole position in a process-independent manner. The infinitely-narrow width approximation of the results is discussed. 

\end{abstract}

\vspace{1cm}
\footnotesize{\begin{tabular}{ll}
{\bf{Pacs:}}$\!\!\!\!$& 12.38.Gc, 13.40.Hq, 13.75.Gx
\\
{\bf{Keywords:}}$\!\!\!\!$& Lattice QCD, transition form factors,
non-relativistic EFT, L\"uscher equation
\\
\end{tabular}}
\end{center}
\end{titlepage}

\setcounter{page}{2}
\section{Introduction}
\label{sec:intro}
Rare $B$ decay modes  provide one of the best opportunities in the search for 
physics beyond the Standard Model (BSM). Among them,  $B\rightarrow K^*l^+l^-$ is regarded as one of the most important channels, as the polarization of the $K^*$ allows a precise angular reconstruction resulting in many observables which can
 be tested in the Standard Model (SM) and its extensions \cite{Ali:1999mm,Bobeth:2012vn,Melikhov:1998cd,Kruger:2005ep,Altmannshofer:2008dz,Matias:2012xw}.
In 2013, LHCb \cite{Aaij:2013qta} published the first analysis of a set of optimized observables, presenting an interesting
pattern of deviations, confirmed by later measurements with a larger
statistics \cite{Aaij:2015oid}, as well as by a recent analysis from the Belle
collaboration \cite{Abdesselam:2016llu}. A first interpretation of this
pattern of deviation was proposed \cite{Descotes-Genon:2013wba}, where the Wilson coefficient $C_9$ of the pertinent semileptonic operator (and, possibly, other
coefficients as well), received contribution from the BSM 
physics. Further experimental
results have indicated deviations concerning the branching ratios of $B\to
K^*\mu^+\mu^-$, but also $B_s\to\phi\mu^+\mu^-$ and $B\to K\mu^+\mu^-$, with the
possibility of a violation of lepton flavour universality between electron
and muon modes~\cite{Aaij:2014pli,Aaij:2014ora,Aaij:2015dea}. These results triggered lots of 
activities on the theoretical side and, in particular, their consequences on global 
fits are being studied \cite{Altmannshofer:2014rta,Hurth:2016fbr,Descotes-Genon:2015uva}. In these global fits, 
a special 
attention has to be paid to
the theoretical uncertainties arising from the {\it form factors} of the corresponding hadronic matrix 
elements,  which affect the branching ratios involved in the
fit. 
In the low recoil region, which will be our main focus here, these form 
factors are  mostly 
known from light cone sum rules, which suffer from relatively large uncertainties \cite{Ball:2004rg,Straub:2015ica}. It would thus be particularly interesting
to have information on these quantities from  lattice QCD simulations. Also, the 
method used
to calculate these form factors could be applied to other
interesting processes as, for example, $B\rightarrow K^*\gamma$.

Recently, the first unquenched lattice QCD calculations of the $B\rightarrow K^*$ form factors have appeared \cite{Liu:2011raa,Horgan:2013hoa,Horgan:2015vla} (see also Refs.~\cite{Bowler:1993rz,Bernard:1993yt,Burford:1995fc,Abada:1995fa,Becirevic:2006nm,Abada:2002ie,Bowler:2004zb} for quenched results). Although this work represents a major progress in the field, the simulations have been performed at such quark mass values that the $K^*(892)$ resonance has been treated as a stable particle. Correspondingly, the standard methods of the lattice QCD could be used for the analysis of the data. However, they are not applicable anymore, when the $K^*$ eventually decays into $\pi K$. 

The following question has to be addressed: how to compute the matrix elements involving two strongly interacting particles in the in- or out-state? Briefly, the answer is given by the so-called Lellouch-L\"uscher method \cite{Lellouch:2000pv}. It is a generalization of the L\"uscher finite-volume approach \cite{Luscher:1990ux},  which provides a method to extract the elastic phase shifts and the resonance parameters (the mass and width) from the two-particle discrete energy levels spectrum, measured on the lattice.

At the next step, it should be understood, how to {\it define} the matrix elements involving resonances such as $K^*,\rho,$ or $\Delta$. As it has been argued in Refs.~\cite{Agadjanov:2014kha,Bernard:2012bi}, the only plausible field-theoretical definition necessitates an analytic continuation of the matrix element to the resonance pole position in the complex plane. Therefore, strictly speaking, the corresponding form factor can only be defined  at the resonance pole. The other well known definition of the form factor is based on the Breit-Wigner parameterization of the resonant amplitude (see, e.g., Refs.~\cite{Aznauryan,Drechsel}). However, this definition yields a model- and process-dependent result, since the background is unknown.
If the width of the resonance is not very small (it is roughly 50 MeV in the case of the 
$K^*(892)$), using different definitions might have an effect on the extracted observables.    

There is an additional effect, which is due to the presence of the $\eta K$ threshold. For physical quark masses, it is approximately 150 MeV above  the $K^*$ mass, and this value will be reduced when the light quark masses, used in the simulations, are higher. One could expect that the effect of this threshold might be seen in the data. The recent lattice calculation by the Hadron Spectrum Collaboration, however, indicates that the  coupling between the $\eta K$ and $\pi K$ channels remains small even at the pion mass as large as roughly 400 MeV \cite{Dudek:2014qha,Wilson:2014cna}. Nevertheless, the two-channel problem has to be addressed. Although of academic interest in the present context, a similar theoretical framework could be useful, e.g., for the lattice extraction of the electromagnetic form factors of the $\Lambda(1405)$ resonance (see Refs.~\cite{Menadue:2013xqa,Hall:2014uca} for the recent lattice results). 

Recently, the Lellouch-L\"uscher method has been generalized to include multiple strongly-coupled decay channels \cite{Hansen:2012tf,Briceno:2012yi,Briceno:2014uqa,Briceno:2015csa}. In particular, the authors of Ref.~\cite{Briceno:2014uqa} provide general formulas for spinless particles, which  are also valid for the $B\rightarrow \pi K (\eta K)$ transition. On the contrary, the extraction of the form factors at the resonance pole in the multi-channel case has not been studied  yet.  It has been done only in the one-channel problem \cite{Agadjanov:2014kha}. In the present work, we fill this gap by considering the $\pi K-\eta K$ coupled-channel system. 

In order to establish a relation between the finite volume quantities, measured on the lattice, and infinite volume observables, a systematic theoretical framework is needed. We apply the so-called non-relativistic effective field theory in a finite volume in its covariant formulation \cite{Colangelo:2006va,Gasser:2011ju}. We find this approach algebraically simpler than the one based on the Bethe-Salpeter equation (see, e.g., Refs.~\cite{Kim:2005gf,Christ:2005gi}). In the end, both methods have the same range of applicability and one arrives at the same results. 

The paper is organized as follows: In section~2, we introduce form factors governing the $B\rightarrow K^*$ transition. We also consider the proper kinematics, which should be used in lattice measurements of matrix elements. Further, in section~3, we set up the non-relativistic effective field theory in a finite volume. The two-channel analogue of the Lellouch-L\"uscher formula is re-derived. In section~4, we obtain the equation for the extraction of the form factors at the resonance pole in the two-channel case. Additionally, in view of different
opinions expressed in the literature (see, e.g., Refs.~\cite{Briceno:2015dca,Briceno:2016kkp}), we 
address the issue of defining the photon virtuality at the resonance pole. In section~5, we consider the infinitely small width approximation for our results. Section~6 contains our conclusions.

\section{Matrix elements on the lattice}
\subsection{Formalism}
The effective theory of the $b\rightarrow s$ decays is based on the weak Hamiltonian \cite{Grinstein:1987vj,Grinstein:1990tj,Buras:1993xp,Ciuchini:1993ks,Ciuchini:1993fk,Ciuchini:1994xa} 
\begin{equation}
\mathcal{H}_{\mathrm{eff}} ~=~ -\frac{4 G_F}{\sqrt{2}}
V_{ts}^* V_{tb}^{} \sum_i C_i W_i \,,
\label{eq:Heff}
\end{equation}
where $G_F$ denotes the Fermi constant, $V_{ts},\, V_{tb}$ are elements of the CKM matrix and the
$C_i$ are Wilson coefficients. In the SM, one has 10 effective local operators $W_i$. Such a description is applicable at energies much below the masses of the weak gauge bosons.

The seven $B\rightarrow K^*$ form factors are contained in the matrix elements of the $W_7,\,W_9$ and $W_{10}$ operators:
\begin{eqnarray}
W_7 = \frac{m_b e}{16\pi^2} \bar{s}\sigma^{\mu\nu} P_R b \, F_{\mu\nu},\quad
W_9 = \frac{e^2}{16\pi^2}\bar{s} \gamma^\mu P_L b\, \bar\ell \gamma_\mu \ell,\quad
W_{10} = \frac{e^2}{16\pi^2}\bar{s} \gamma^\mu P_L b\, \bar\ell \gamma_\mu 
\gamma^5 \ell,
\end{eqnarray}
where $F_{\mu\nu}$ is the electromagnetic field strength tensor, and 
\begin{equation}
P_{L/R} = \tfrac{1}{2}(1 \mp \gamma^5),\quad \sigma^{\mu\nu} =
\tfrac{i}{2}[\gamma^\mu,\gamma^\nu].
\end{equation}
They are defined, in Minkowski space, through the following expressions (see, e.g., Ref.~\cite{Horgan:2013hoa}):
\begin{eqnarray}\label{formfact1}
\langle V(k,\lambda) | \bar{s}\gamma^\mu b| B(p)\rangle
&=& \frac{2 i  V(q^2)}{m_B + m_V}
\epsilon^{\mu\nu\rho\sigma} \epsilon^*_\nu k_\rho p_\sigma, \\[4mm]
\langle V(k,\lambda)|\bar{s}\gamma^\mu \gamma^5 b| B(p)\rangle
&=&  2 m_V A_0(q^2) \frac{\epsilon^* \cdot q}{q^2} q^\mu 
 + (m_B + m_V)A_1(q^2)\left(
\epsilon^{*\mu} - \frac{\epsilon^* \cdot q}{q^2} q^\mu \right)\nonumber \\ 
&&- \; A_2(q^2) \frac{\epsilon^*\cdot q}{m_B + m_V}\left[
(p+k)^\mu - \frac{m_B^2 - m_V^2}{q^2} q^\mu\right], \\[4mm]
q_\nu\langle V(k,\lambda) | \bar{s}\sigma^{\mu\nu}b| B(p)\rangle
&=&  2 T_1(q^2) \epsilon^{\mu\rho\tau\sigma} \epsilon^*_\rho p_\tau k_\sigma, 
\\[4mm]
q_\nu\langle V(k,\lambda) | 
\bar{s}\sigma^{\mu\nu} \gamma^5 b| B(p)\rangle
& = &  i  T_2(q^2)
[(\epsilon^* \cdot q)(p+k)^\mu - \epsilon^{*\mu} (m_B^2-m_V^2) 
] \nonumber \\\label{formfact4}
&&+\; i T_3(q^2)(\epsilon^*
\cdot q)\left[ \frac{q^2}{m_B^2
-m_V^2}(p+k)^\mu - q^\mu \right]\,,
\end{eqnarray}
where $q=p-k$ is a momentum transfer to the lepton pair, and $\epsilon(k,\lambda)$ denotes a polarization vector of the vector meson ($K^*$) with momentum $k$ and spin polarization $\lambda=1,2,3$ (see, e.g., Ref.~\cite{Horgan:2013hoa}). Here, it is assumed that the $K^*$ is a stable particle with mass $m_V$ and appropriate quantum numbers. 

We note that the contributions of other operators to the full decay amplitude are seen to be small in the low recoil region, in which both $B$ and $K^*$ are roughly at rest (see, e.g., Refs.~\cite{Khodjamirian:1997tg,Khodjamirian:2010vf,Grinstein:2004vb,Beylich:2011aq}). Correspondingly, we consider the decay process in this kinematic region, so that the amplitude, extracted from lattice data, coincides approximately with the full one.

\subsection{Finite volume}
Since lattice simulations are performed in a finite spatial volume, the continuous rotational symmetry is broken down to the cubic one. Consequently, some particular irreducible representations (irreps) of the cubic group, or its subgroups in the moving frames, should be chosen. Taking into account the fact that,
at energies below multi-particle  thresholds the neglect  of D- and 
higher partial waves seems to be justified, in order to clearly 
extract  the P-wave scattering phase shift through the L\"uscher equation, 
it is preferable 
to choose irreps, in which  no mixing between S- and P- waves occurs. 
For that purpose, we consider the process in the $K^*$ rest frame: 
\begin{equation}
{\bf k}=0,\quad {\bf p}={\bf q}=\frac{2\pi}{L}{\bf d}, \quad {\bf d} \in \mathbb{Z}^3,
\end{equation}
where $L$ denotes the side length of the volume, $V=L^3$.
When the $K^*$ is not at rest, only some of the form factors can be extracted without mixing. We provide the details in Appendix A. In the following, we write down the expressions for the current matrix elements, when the ${\bf d}$ vector is chosen along the third axis ${\bf d}=(0,0,n)$. The two other cases, ${\bf d}=(n,n,0)$ and ${\bf d}=(n,n,n)$ can be treated along the same lines.

The polarization vector of the free massive spin-1 particle with momentum ${\bf k}$ takes the form:
\begin{equation} 
\epsilon^\mu(k,\lambda)=\biggl(\frac{{\bf k}\cdot{\boldsymbol \epsilon}^{(\lambda)}}{m_V},\,{\boldsymbol \epsilon}^{(\lambda)}+\frac{{\bf k}\cdot{\boldsymbol \epsilon}^{(\lambda)}}{m_V(k_0+m_V)}{\bf k} \biggr), 
\end{equation}  
where the arbitrary vectors ${\boldsymbol \epsilon}^{(\lambda)}$ form an orthonormal basis. In particular, one can choose them as
\begin{equation}
{\boldsymbol \epsilon}^{(+)}=\frac{1}{\sqrt{2}}(1,i,0),\quad{\boldsymbol \epsilon}^{(-)}=\frac{1}{\sqrt{2}}(1,-i,0),\quad{\boldsymbol \epsilon}^{(0)}=(0,0,1).
\end{equation}
Obviously, the polarization vectors $\epsilon^\mu(k,\lambda)$ satisfy the gauge invariance condition
\begin{equation}
k_\mu\cdot\epsilon^\mu(k,\lambda)=0,\quad \lambda=+,-,0.
\end{equation}

Further, the Eqs. (\ref{formfact1})-(\ref{formfact4}) first have to be rewritten in the Euclidean space. This can be done by applying the prescription
\begin{equation}
a^E_\mu=({\bf a},ia_0),\quad \gamma^E_\mu=(-i{\boldsymbol \gamma},\gamma_0),\quad \gamma^E_5=\gamma^5,\qquad \mu=1,2,3,4,
\end{equation}
where $a^\mu$ is an arbitrary four-momentum in Minkowski space. The superscript $E$ will be suppressed from now on. 
\begin{table}[placement]
\begin{center}
\begin{tabular}{|c|c|c|}
\hline Little group & Irrep &  Form factor\\ [1ex]
\hline\hline \multirow{2}{*}{$C_{4v}$} & $\mathbb{E}$ & $V$,\,$A_1$,\,$T_1$,\,$T_2$ \\[1ex] 
   & $\mathbb{A}_1$ & $A_0$,\,$A_{12}$,\,$T_{23}$\\ 
   
\hline 
\end{tabular} 
\end{center}
\caption{\small Extraction of matrix elements in the irreps without partial-wave mixing.}
\end{table}

With this in mind, we pick up the following current matrix elements 
\begin{eqnarray}\nonumber\label{7ffs}
\langle V(+) | J^{(+)}| B(p)\rangle
&=& -\frac{2 i m_V|{\bf q}|V(q^2)}{m_B + m_V},\\[2mm]
\langle V(0) | i(E_B-m_V) J_A + |{\bf q}| J^{(0)}_A | B(p)\rangle
&=& - 2im_V|{\bf q}| A_0(q^2),\nonumber\\[2mm]
\langle V(+)|J^{(+)}_A| B(p)\rangle
&=& -i(m_B+m_V) A_1(q^2),\nonumber\\[2mm]
\langle V(0) |   i(E_B-m_V)J^{(0)}_A-|{\bf q}| J_A | B(p)\rangle
&=&  8m_Bm_V A_{12}(q^2),\nonumber\\[2mm]
\langle V(+) | i(E_B-m_V)I^{(+)} + |{\bf q}|I^{(+)}_0 | B(p)\rangle
&=&   2im_V|{\bf q}| T_1(q^2),\nonumber\\[2mm]
\langle V(+) | i(E_B-m_V)I^{(+)}_{A} + |{\bf q}|I^{(+)}_{0A} | B(p)\rangle
&=&  -i(m_B^2-m_V^2) T_2(q^2),\nonumber\\[2mm]
\langle V(0) | 
I^{(0)}_A| B(p)\rangle
&=&- \frac{4m_Bm_V}{m_B+m_V} T_{23}(q^2),
\end{eqnarray}
where $E_B=\sqrt{m^2_B+{\bf q}^2}$ is energy of the $B$ meson, and  $\langle V(+)|$ is a state vector with a positive circular polarization,
\begin{equation}
\langle V(+)|=\frac{\langle V(1)|-i\langle V(2)|}{\sqrt{2}}~.
\end{equation}
Here, the  current operators are given by
\begin{eqnarray}\nonumber
J^{(\pm)}&=&\frac{1}{\sqrt{2}}\bar{s}(\gamma_1\pm i\gamma_2) b, \quad J^{(\pm)}_A=\frac{1}{\sqrt{2}}\bar{s}(\gamma_1\pm i\gamma_2) \gamma_5 b,
\\[2mm]\nonumber
J^{(0)}_A&=&\bar s \gamma_3\gamma_5 b,\quad J_A=\bar s \gamma_4\gamma_5 b,\quad I^{(0)}_A=\bar{s}\sigma_{34} \gamma_5 b,
\\[2mm]\nonumber
I^{(\pm)}_0&=&\frac{1}{\sqrt{2}}\bar{s}(\sigma_{13}\pm i\sigma_{23})b,\quad I^{(\pm)}_{0A}=\frac{1}{\sqrt{2}}\bar{s}(\sigma_{13}\pm i\sigma_{23})\gamma_5b,
\\[2mm]
 I^{(\pm)}&=&\frac{1}{\sqrt{2}}\bar{s}(\sigma_{14}\pm i\sigma_{24})b,\quad I^{(\pm)}_{A}=\frac{1}{\sqrt{2}}\bar{s}(\sigma_{14}\pm i\sigma_{24})\gamma_5b,
\end{eqnarray}
and the quantities $A_{12}(q^2)$, $T_{23}(q^2)$ are related to the form factors through
\begin{align}
A_{12}(q^2) \;=\; & \frac{(m_B + m_V)^2(m_B^2 - m_V^2 - q^2) A_1(q^2) -
\lambda A_2(q^2)}{16 m_B m_V^2 (m_B + m_V)}, \\
T_{23}(q^2) \;=\; & \frac{m_B+m_V}{8m_B m_V^2}
\left[\left(m_B^2 + 3m_V^2 - q^2\right)T_2(q^2)
- \frac{\lambda T_3(q^2)}{m_B^2 - m_V^2} \right], 
\end{align}
where $\lambda\equiv\lambda(m_B^2,m_V^2, q^2)=[(m_B+m_V)^2-q^2][(m_B-m_V)^2-q^2]$ denotes the K\"all\'en triangle function. In the following, we denote the matrix elements Eq.~(\ref{7ffs}) shortly as $F^M$, $M=1,...,7$.

When the $K^*$ is taken at rest, it is necessary to consider lattice simulations 
in asymmetric boxes (see below). These boxes, which are of the type 
$L\times L\times L'$, have the same symmetry properties as the symmetric ones 
boosted in the ${\bf d}=(0,0,n)$ direction. In Table~1, 
the irreps of the corresponding little group, where the matrix elements 
Eq.~(\ref{7ffs}) should be measured, are listed. 

The states $\langle V(\pm) |,\,\langle V(0) |$ are created by acting with
the following local field operators, transforming according to these irreps, on the vacuum state $\langle 0 |$:
\begin{eqnarray}\label{operators}
{\cal O}_{\mathbb{E}}^{(\pm)}({\bf 0},t)=\frac{1}{\sqrt{2}}\sum_{\bf x}\big(O_1({\bf x},t)\mp iO_2({\bf x},t)\big),\quad {\cal O}_{\mathbb{A}_1}^{(0)}({\bf 0},t)=\sum_{\bf x}O_3({\bf x},t),
\end{eqnarray}
where $O_i(x)$ are spatial components of the vector field potential (see, e.g., Ref.~\cite{Gockeler:2012yj}). Such operators are constructed out of the local quark bilinears. In practice, it is important to add also meson-meson-type non-local operators in lattice simulations. These can be constructed along the lines described in Ref.~\cite{Gockeler:2012yj}.

Until now, the $K^*$ has been assumed to be a stable vector meson. When the $K^*$ becomes a resonance in lattice simulations, the matrix elements of Eq.~(\ref{7ffs}) can still be measured. However, one gets the matrix elements of the current between  a one-meson state $|B(p)\rangle$ and a certain eigenstate of the finite-volume  Hamiltonian. The mass $m_V$ is now replaced by the discrete energy $E_n$ of the $n$-th eigenstate ($n=0,1,...$). The dependence of the energy $E_n$ on the volume is not suppressed exponentially (unlike the case of a stable $K^*$) \cite{Luscher:1990ux}. A similar statement holds for the quantities $F^M$.

The matrix elements $F^M$ are functions of the total center-of mass (CM) energy $E_n$ and 3-momentum $|{\bf q}|$ of the $B$ meson: $F^M=F^M(E_n,|{\bf q}|)$. As it has been previously discussed in case of the $\Delta N\gamma^*$ transition in Ref.~\cite{Agadjanov:2014kha}, in order to determine the form factors at the $K^*$ resonance pole, the quantities $F^M$ should be measured at different values of the energy $E_n$ (for a given value of $n$), while keeping $|{\bf q}|$ fixed. Again, this could be achieved by applying asymmetric volumes with asymmetry along the third axes $L\times L\times L'$  or (partial) twisting in the $b$-quark (see Ref.~\cite{Agadjanov:2014kha} for more details).

Below, we study in detail the extraction of the form factors on the real energy axis as well as at the complex resonance pole. We emphasize once more, that only the definition, which implies the analytic continuation, leads to the process-independent values of the resonance form factors.

\section{Lellouch-L\"uscher formula}
\subsection{Infinite volume}
In this section, the analogue of the Lellouch-L\"uscher formula in the two-channel case is reproduced. For that purpose, we apply the non-relativistic effective field theory in a finite volume along the lines of Refs.~\cite{Bernard:2012bi,Agadjanov:2014kha}. We generalize the formulas given there appropriately  so that they can suit our needs. In the following, the $K^*$ is taken at rest, so that there is no S- and P-wave mixing. 

Further, we specify the matrix elements of the scattering amplitude. The actual  physics can not, of course, depend on the chosen parameterization. In the literature,
 there exists a  parameterization of the $S$-matrix due to 
Stapp {\it et al.}~\cite{Stapp:1956mz}. 
In this work, we rather follow the one from 
Refs.~\cite{Hansen:2012tf,Blatt:1952zza} and write the 
$T$-matrix in terms of three real parameters: the so-called eigenphases 
$\delta_1(p_1)$, $\delta_2(p_2)$ and mixing parameter $\varepsilon(E)$
\begin{equation}\label{T-matrix}
T={8\pi\sqrt{s}}
\begin{pmatrix}
\frac{1}{p_1}(\ce^2e^{i\df}\sin\df+\se^2e^{i\ds}\sin\ds) &\frac{1}{\sqrt{p_1p_2}}\ce\se(e^{i\df}\sin\df-e^{i\ds}\sin\ds)\\ 

\frac{1}{\sqrt{p_1p_2}}\ce\se(e^{i\df}\sin\df-e^{i\ds}\sin\ds) & \frac{1}{p_2}(\ce^2e^{i\ds}\sin\ds+\se^2e^{i\df}\sin\df)
\end{pmatrix},
\end{equation}
where $\se\equiv\sin\varepsilon(E)$, $\ce\equiv\cos\varepsilon(E)$. Here, $p_1$ and $p_2$ denote the relative 3-momenta in the $\pi K$ and $\eta K$ channels, respectively. They are related to the total energy $E$ through the equations
\begin{equation}\label{3-momenta}
|p_1|=\frac{\lambda(m_\pi^2,m_K^2, s)}{2\sqrt{s}},\quad |p_2|=\frac{\lambda(m_\eta^2,m_K^2, s)}{2\sqrt{s}},
\end{equation}
where $s=E^2$. We note that the eigenphases $\delta_1$, $\delta_2$ have the meaning of phase shifts in the corresponding channels $\pi K$ and $\eta K$, respectively, only in the decoupling limit $\varepsilon\rightarrow 0$. Otherwise, their behaviour with energy is non-trivial (see, e.g., Refs.~\cite{Dalitz:1970,Workman:2012hx}). 
Firstly, thanks to the no-crossing theorem \cite{Wigner:1929}, the curves of the functions $\delta_1(E)$, $\delta_2(E)$ cannot intersect. Secondly, assuming the Breit-Wigner approximation, it can be shown that only one of these curves crosses $\pi/2$ in the vicinity of the resonance energy (see below). Lattice data should not be in contradiction with
 these properties.

On the other hand, the $T$-matrix obeys Lippmann-Schwinger equation (see Ref.~\cite{Bernard:2012bi}):
\begin{equation}\label{LS}
T=V+VGT,
\end{equation} 
where the angular momentum index $l$ has been suppressed. Here, $V$ denotes a potential and $G(s)$ is a loop function matrix given by 
\begin{equation}\label{loop function}
G=
\begin{pmatrix}
\frac{ip_1}{8\pi\sqrt{s}} & 0\\
0 & \frac{ip_2}{8\pi\sqrt{s}}
\end{pmatrix}.
\end{equation} 
In Eq.~(\ref{LS}), all quantities  have been taken on the energy shell $p_1=p'_1,\,p_2=p'_2$, 
where $p_1,\,p_2$ and $p'_1,\,p'_2$ are respective relative momenta in the initial and final 
two-particle states.

The parameterization of the potential $V$ in terms of parameters $\delta_1(p_1)$, $\delta_2(p_2)$ and $\varepsilon(E)$ is obtained readily  from Eqs.~(\ref{T-matrix}) and~(\ref{LS}):

\begin{equation}\label{potential}
V={8\pi\sqrt{s}}
\begin{pmatrix}
\frac{1}{p_1}(t_1+\se^2t) &-\frac{1}{\sqrt{p_1p_2}}\ce\se t\\ 

-\frac{1}{\sqrt{p_1p_2}}\ce\se t & \frac{1}{p_2}(t_2-\se^2t)
\end{pmatrix},
\end{equation}
where $t_i\equiv\tan\delta_i(p_i)$ and $t=t_2-t_1$. Clearly, the potential matrix $V$ is real and symmetric. 

\subsection{Finite volume}
\subsubsection{Two-point function}
We return to the derivation of the two-channel Lellouch-L\"uscher formula. Our goal is to calculate the two- and three-point correlation functions relevant to the $B\rightarrow K^*$ form factors. Let $O(x)$ be a local operator with quantum numbers of the $K^*$ that transforms according to the given irrep, as provided explicitly in Eq.~(\ref{operators}). According to the methodology of the lattice calculations, one is interested in the Euclidean two-point function of the form 
\begin{equation}\label{two-point}
D(x_0-y_0)=\langle0|{\cal O}(x_0){\cal O}^{\dagger}(y_0)|0\rangle,
\end{equation}
where ${\cal O}(t)$ is given by the Fourier transformation of the $O(x)$ in the rest frame:
\begin{equation}\label{O-def}
{\cal O}(t)=\sum_{{\bf x}}O({\bf x},t).
\end{equation}
Note that we always work in the limit of zero lattice spacing, in which the right-hand side of Eq. (\ref{O-def}) contains an integral over the finite volume instead of a sum over the lattice sites.
 
It is clear from the spectral representation\footnote{In this work, we use a different from Ref.~\cite{Agadjanov:2014kha} normalization of the eigenstates of the total Hamiltonian. While the single $B$-meson state in a finite
volume is still normalized, according to
$\langle B({p})|B({p})\rangle=2E_B$, the normalization of the two-particle states $E_n$ is given by $\langle E_n|E_n\rangle=1$.} of the function $D(x_0-y_0)$,
\begin{equation}\label{2-point-spectral}
D(x_0-y_0)=\sum_{{n}}e^{-E_n(x_0-y_0)}|\langle 0|{\cal O}(0)|E_n\rangle|^2,
\end{equation}
that energy levels $E_n$ can be extracted by studying the decay pattern of $D(x_0-y_0)$ in the formal limit $x_0\rightarrow+\infty,\, y_0\rightarrow-\infty$.

The diagrammatic representation of the two-point function Eq. (\ref{two-point}) within the non-relativistic effective field theory below the inelastic threshold is shown in Fig.~1. The quantities $X_\alpha,\,\alpha=1,2$, denote the couplings of the operator ${\cal O}$ to respective channels. Since the corresponding Lagrangian contains terms with arbitrary number of spatial derivatives, one has $X_\alpha=A_\alpha+B_\alpha {\bf p}^2_\alpha+\cdots$, where $A_\alpha,B_\alpha,\dots$ contain only short-range physics. Here, ${\bf p}^2_\alpha,\, \alpha=1,2$, are {\it external} relative 3-momenta squared in the corresponding channels. Although the expansion for $X_\alpha$ is written in the CM frame, it can be brought to the covariant form in an arbitrary moving frame (see Ref.~\cite{Gasser:2011ju}). It is important to note that quantities $X_\alpha$ will drop out in the final result. 
\begin{figure}[placement]
\begin{center}
\includegraphics[scale=0.55]{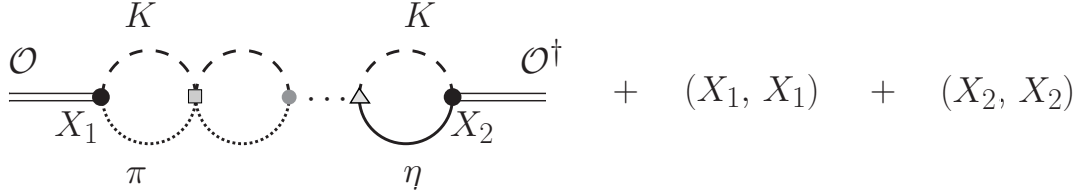}
\end{center}
\caption{\small Two-point function $D(x_0-y_0)$ in the non-relativistic effective field theory in a finite volume. The grey circle, square, and triangle depict different couplings in the $\pi K-\eta K$ system. The quantities $X_1,X_2$ are couplings of the operator ${\cal O}$ to the respective channels.  Similar diagramms are obtained by replacements $X_1\to X_2$ and $X_2\to X_1$.}
\end{figure}

After summing up all two-particle reducible diagrams, the two-point function reads

\begin{equation}\label{2point-integral}
D(x_0-y_0)={\cal V}\int_{-\infty}^{+\infty}\frac{dP_0}{2\pi}\,e^{iP_0(x_0-y_0)}X^T[G_L(P_0)+G_L(P_0) T_L(P_0) G_L(P_0)]X,
\end{equation}
where $X^T=(X_1,X_2)$, ${\cal V}$ is the lattice volume,  and $G_L$ denotes a finite-volume counterpart of the loop function matrix Eq.~(\ref{loop function}):
\begin{equation}
G_L=
\begin{pmatrix}
-\frac{p_1}{8\pi\sqrt{s}}\cot\phi(p_1)\ & 0\\
0 & -\frac{p_2}{8\pi\sqrt{s}}\cot\phi(p_2)
\end{pmatrix},\quad s=-P_0^2.
\end{equation} 
Here, $\phi(p_\alpha)$ are the volume-dependent functions that are related to the L\"uscher zeta-function. They are given by the following expressions in the irreps of interest $\mathbb{E}$ and $\mathbb{A}_1$  (see, e.g., Ref.~\cite{Gockeler:2012yj}):
\begin{eqnarray}
\cot\phi^\mathbb{E}(p_\alpha)=-\frac{1}{{\pi}^{3/2}\eta_\alpha}\,\biggl\{\hat Z_{00}(1;\eta_\alpha^2)-\frac{1}{\sqrt{5}\eta_\alpha^2}\,
\hat Z_{20}(1;\eta_\alpha^2)\biggr\},\\
\cot\phi^{\mathbb{A}_1}(p_\alpha)=-\frac{1}{{\pi}^{3/2}\eta_\alpha}\,\biggl\{\hat Z_{00}(1;\eta^2_\alpha)+\frac{2}{\sqrt{5}\eta^2_\alpha}\,
\hat Z_{20}(1;\eta_\alpha^2)\biggr\},
\end{eqnarray}
where $\eta_\alpha=p_\alpha L/2\pi$. The L\"uscher zeta-function $\hat Z_{lm}(1;\eta^2)$ for generic asymmetric volumes $L\times L\times L'$ with $L'=xL$ reads 
\begin{equation}
\hat Z_{lm}(1;\eta^2)=\frac{1}{x}\sum_{\bf n\in\mathbb{Z}^3}\frac{{\cal Y}_{lm}({\bf r})}
{{\bf r}^2-\eta^2}\, ,\quad\quad r_{1,2}=n_{1,2}\, ,\quad r_3=\frac{1}{x}\,n_3\,
,\quad\quad \hat Z_{20}(1;\eta^2)\neq 0\, .
\end{equation}
Further, the $T_L$-matrix is a scattering amplitude in a finite volume that is defined formally also through a Lippmann-Schwinger equation with the same potential $V$:
\begin{equation}
T_L=V+VG_LT_L.
\end{equation}
Substituting the potential $V$, Eq.~(\ref{potential}), into this equation, we obtain:
\begin{equation}\label{2T_L}
T_L=\frac{8\pi\sqrt{s}}{f(E)}
\begin{pmatrix}
\frac{1}{p_1}[t_1\tf(t_2+\ts)+\se^2\tf\ts t] &-\frac{1}{\sqrt{p_1p_2}}\ce\se \tf\ts t\\ 

-\frac{1}{\sqrt{p_1p_2}}\ce\se \tf\ts t& \frac{1}{p_2}[t_2\ts(t_1+\tf)-\se^2\tf\ts t]
\end{pmatrix},
\end{equation}
where $\tau_\alpha\equiv\tan\phi(p_\alpha)$ and 
\begin{equation}\label{L-equation}
f(E)\equiv(t_1+\tau_1)(t_2+\tau_2)+\se^2(t_2-t_1)(\tau_2-\tau_1)\textsc{\textsc{\textsc{\textsc{}}}}.
\end{equation}
The two-channel L\"uscher equation  \cite{Hansen:2012tf,He:2005ey,Liu:2005kr}, which allows to determine the infinite-volume $T$-matrix elements ~\cite{Hansen:2012tf,Bernard:2010fp,Doring:2011vk}, follows directly from Eq.~(\ref{L-equation}) 
\begin{equation}
(t_1+\tau_1)(t_2+\tau_2)+\se^2(t_2-t_1)(\tau_2-\tau_1)\big|_{E=E_n}=0,
\end{equation}
where all quantities are taken at 
the energies $E=E_n$ of the simple poles of the $T_L$-matrix, or equivalently, the eigenvalues of the corresponding strong Hamiltonian in a finite volume.

The integral Eq.~(\ref{2point-integral}) is evaluated by applying Cauchy's theorem. It can be shown explicitly that only the poles of the $T_L(P_0)$-matrix contribute to the integral, while free poles cancel in the integrand \cite{Bernard:2012bi,Briceno:2014uqa}. The residues of the $T_L(P_0)$ factorize in the $n$-th pole $P_0=iE_n$:
\begin{equation}
T_L^{\alpha\beta}=\frac{f_\alpha f_\beta}{E_n+iP_0}+\cdots.
\end{equation}
Here, the quantities $f_1,\,f_2$ can be brought to the following form by applying the L\"uscher equation:
\begin{eqnarray}
f_1^2=\frac{8\pi\sqrt{s}}{p_1}\frac{\tf^2(t_2+\ts-\se^2 t)}{f'(E)}\bigg|_{E=E_n},\quad
f_2^2=\frac{8\pi\sqrt{s}}{p_2}\frac{\ts^2(t_1+\tf+\se^2 t)}{f'(E)}\bigg|_{E=E_n},
\end{eqnarray}
where $f'(E)\equiv{df(E)}/{dE}$. Performing the integration over $P_0$, we get
\begin{equation}
D(x_0-y_0)=\frac{{\cal V}}{64\pi^2E_n^2}\sum_{{n}}e^{-E_n(x_0-y_0)} \,\bigg[\sum_{\alpha=1}^{2} X_\alpha p_\alpha(E_n)\tau_\alpha^{-1} (E_n)f_\alpha(E_n)\bigg]^2.
\end{equation}
Comparing this equation with the spectral representation Eq.~(\ref{2-point-spectral}), we finally obtain
\begin{equation}\label{2point-final}
|\langle 0|{\cal O}(0)|E_n\rangle|=\frac{{\cal V}^{1/2}}{8\pi E_n}\bigg|\sum_{\alpha=1}^{2} X_\alpha p_\alpha(E_n)\tau_\alpha^{-1} (E_n)f_\alpha(E_n)\bigg|.
\end{equation}

\subsubsection{Three-point function}
We proceed to evaluate the current matrix elements $F^M(E,|{\bf q}|)$ in a finite volume. To this end, we start from the quantity
\begin{equation}
\Gamma^M(x_0,p)=\langle 0|{\cal O}(x_0)J^M(0)|B(p)\rangle,\quad M=1,\dots 7.
\end{equation}
Here, the $J^M(0)$ denote the operators in the matrix elements of Eq. (\ref{7ffs}). Inserting
a complete set of states, we get the spectral representation of $\Gamma^M(x_0,p)$
\begin{equation}\label{3point-spectral}
\Gamma^M(x_0,p)=\sum_ne^{-E_nx_0}\langle 0|{\cal O}(0)|E_n\rangle F^M(E_n,|{\bf q}|).
\end{equation}

\begin{figure}[placement]
\begin{center}
\includegraphics[scale=0.38]{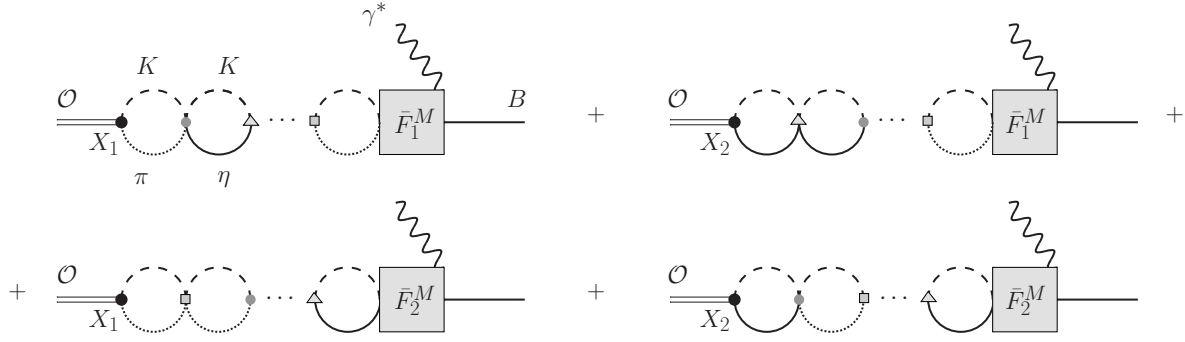}
\end{center}
\caption{\small Diagrams contributing to the $B\to K^*$ transition in a finite volume (see Fig.~1 for notations). The quantities $\bar F^M_\alpha(E,|{\bf q}|),\, \alpha=1,2$, are volume-independent up to  exponentially suppressed contributions.}
\end{figure}

Diagrammatically, the $B\rightarrow K^*$ transition matrix elements are shown in Fig.~2. The quantities $\bar F^M_\alpha(E,|{\bf q}|),\, \alpha=1,2$, denote the sum of all two-particle irreducible diagrams in the respective channels. They 
do not depend on the volume up to exponentially suppressed contributions. The volume dependence arises due to the final-state meson interaction. We note that the diagrams, in which the photon is attached to one of the internal lines or the $B$ meson external line do not contribute to the matrix elements of {\it flavor changing neutral currents}. As a result, summing up the bubble diagrams we obtain
\begin{equation}
\label{3point-integral}
\Gamma^M(x_0,p)={\cal V}^{-1/2}\int_{-\infty}^{+\infty}\frac{dP_0}{2\pi}\,e^{iP_0x_0}X^T[G_L(P_0)+G_L(P_0) T_L(P_0) G_L(P_0)]\bar F^M(P_0,|{\bf q}|),
\end{equation} 
where $\bar F^M(P_0,|{\bf q}|)$ denotes a two-component vector with elements $\bar F^M_\alpha(P_0,|{\bf q}|)$. Similarly to the case of the two-point function, only the poles of the $T_L(P_0)$-matrix contribute to the integral. Integrating over $P_0$, one gets
\begin{eqnarray}
\Gamma^M(x_0,p)&=&\frac{{\cal V}^{-1/2}}{64\pi^2E_n^2}\sum_{{n}}e^{-E_nx_0}\sum_{\alpha,\beta=1}^{2} [X_\alpha p_\alpha (E_n)\tau^{-1}_\alpha(E_n)f_\alpha(E_n)]\nonumber\\[0.2cm]&\times&[p_\beta(E_n)\tau^{-1}_\beta(E_n)f_\beta(E_n)\bar F_\beta^M(E_n,|{\bf q}|) ].
\end{eqnarray}
Comparing this formula with Eq.~(\ref{3point-spectral}) and using Eq.~(\ref{2point-final}), we arrive at the final result:
\begin{equation}\label{LL-equation}
|F^M(E_n,|{\bf q}|)|=\frac{{\cal V}^{-1}}{8\pi E}\big|p_1\tau^{-1}_1f_1\,\bar F^M_1+p_2\tau^{-1}_2f_2\,\bar F^M_2\big|\bigg|_{E=E_n}.
\end{equation}

The last step that needs to be done is to relate the above defined quantities $\bar F_1^M,\,\bar F_2^M$ to the (infinite-volume) decay amplitudes ${\cal A}^M_1(B\rightarrow\pi Kl^+l^-)$ and ${\cal A}^M_2(B\rightarrow\eta Kl^+l^-)$ through the two-channel Watson theorem. After summing up the two-particle reducible diagrams in the infinite volume, one gets
\begin{equation}
{\cal A}^M=(1-VG)^{-1}\bar F^M,
\end{equation}
or
\begin{equation}
{\cal A}^M=TV^{-1}\bar F^M,
\end{equation}
where the Lippmann-Schwinger equation has been used. We obtain:
\begin{eqnarray}\label{Watson}
{\cal A}^M_1=\frac{1}{\sqrt{p_1}}(u_1^M\ce e^{i\delta_1}-u_2^M\se e^{i\delta_2}),\quad {\cal A}^M_2=\frac{1}{\sqrt{p_2}}(u_2^M\ce e^{i\delta_2}+u_1^M\se e^{i\delta_1}),
\end{eqnarray}
where
\begin{equation}\label{u1u2}
u_1^M=(\sqrt{p_1}\ce\bar F^M_1+\sqrt{p_2}\se\bar F^M_2)\cos\df,\quad u_2^M=(\sqrt{p_2}\ce\bar F^M_2-\sqrt{p_1}\se\bar F^M_1)\cos\ds.
\end{equation}
We have arrived at the two-channel analog of the Lellouch-L\"uscher formula for the $B\to K^*$ transition. Note that, writing  Eq.~(\ref{LL-equation}) in terms of the amplitudes $u_1^M,\, u_2^M$, one 
obtains the expressions similar to ones given in Ref.~\cite{Hansen:2012tf}. Later, we will consider the limit of this result, when the $K^*$ resonance is infinitely narrow.

Hence, in the two-channel case, two quantities ${\bar F}^M_1$, ${\bar F}^M_2$ and their relative 
sign have to be determined from one equation, whereas in the one-channel case,
 only one quantity for one equation was involved. Consequently, one needs at least three different measurements at the same energy. This involves the extraction of the excited energy levels (see Ref.~\cite{Hansen:2012tf}). An alternative would be to measure the same energy level in asymmetric volumes of type $yL\times yL \times L'$ for different values of parameter $y$ and $L'$ fixed.
Also, as long as one does not insist on keeping the variable $|{\bf q}|$ fixed
and is ready to perform a two-variable fit for the quantities 
${\bar F}^M_\alpha$, (partial) twisting in the $s$-quark or boosts can be applied. 
Then, the spectrum  becomes dependent on the value of the twisting angle and/or the boost 
momentum. Although this option appears to be promising \cite{Doring:2011vk}, the 
(potentially large) S- and P-wave mixing  is inevitable in this case.   

\section{Form factors at the {\boldmath$K^*$} resonance pole}

The current matrix elements involving resonances have the proper field-theoretical meaning only if they are analytically continued to the resonance pole position. The advantage of such a definition is that it is process-independent. On the other hand, the definition based on the Breit-Wigner parameterization is, generally, not free of process- and model-dependent ambiguities, since the non-resonant background is unknown. 

\subsection{Effective-range expansion}
The first step towards the pole extraction of the $B\to K^*$ form factors consists in the determination  of the $K^*$ resonance position. As is well known, the resonances are associated with complex poles of the
scattering amplitude $T$ on unphysical Riemann sheets in the energy plane ($s$ plane). 
The $T$-matrix itself is analytic on the whole plane except for cuts and poles.
Here we will assume that all distant singularities from the pole do not affect
the determination of its position. Thus, 
from the 
analytic structure of the functions $p_1(E),\, p_2(E)$, Eq.~(\ref{3-momenta}), 
the only relevant singularities for our purpose 
are two cuts, which run from branch points at the threshold energies $E_1=m_K+m_\pi$ and $E_2=m_K+m_\eta$, respectively, along the positive axis to infinity. 
The imaginary parts of the $p_\alpha(s),\,\alpha=1,2$, change the sign, when one goes from one sheet to another through these cuts. The four Riemann sheets are classified according to the signs of ${\rm Im}p_1$ and ${\rm Im}p_2.$ (see, e.g., Ref.~\cite{Badalian:1981xj}). For example, on the sheet II one has ${\rm Im}p_1<0$ and ${\rm Im}p_2>0$, etc. 

Further, it is convenient to formulate the problem in the $K$-matrix formalism. The $l=1$ partial-wave amplitude $T$ is defined in terms of $K$-matrix as follows:
\begin{equation}
T=(8\pi\sqrt{s})(K^{-1}-iP )^{-1},
\end{equation}
where $P={\rm diag} (p_1,p_2)$ is a diagonal matrix.
A comparison of this equation with Eq.~(\ref{LS}) leads to the conclusion that  the $K$-matrix is proportional to the potential $V$:
\begin{equation}
K=(8\pi\sqrt{s})^{-1}V.
\end{equation}
The poles of the scattering amplitude $T$ appear as the complex solutions of the secular equation, which we write as
\begin{equation}\label{pole-position}
\det(PK^{-1}P-iP^3)=0.
\end{equation}
The explicit form of this equation is different on each Riemann sheet. For instance, if one is interested in the solutions on the sheets II and  III, then the matrix $P$ must chosen as $P_{II}={\rm diag} (-p_1,p_2)$ and $P_{III}={\rm diag} (-p_1,-p_2)$, respectively. The change of sign of momenta $p_1$ and/or $p_2$ is equivalent to the transition from one sheet to another. 

The analytic properties of the $K$-matrix ensure that the $PK^{-1}P$ function obeys a polynomial expansion of the form (see Refs.~\cite{Badalian:1981xj,Ross:1961})
\begin{equation}\label{ERE}
PK^{-1}P=A+B(E-E_0)+\cdots,
\end{equation}
where $E_0$ is an arbitrary point on the real axis, around which the Taylor expansion is made. The formula Eq.~(\ref{ERE}) is a multi-channel generalization of the well-known effective-range approximation \cite{Bethe:1949yr}. Its additional advantage is the freedom to choose the value of the energy $E_0$: one does not need to start the expansion at threshold energies, as it is usually done. Consequently, the convergence of the series in Eq.~(\ref{ERE}) could be substantially improved. Also, the analytic continuation to the resonance pole position will not be spoiled by the presence of the distant singularities. This expansion, in particular, might be useful in case of the $\rho$ resonance, when lattice simulations are performed at nearly physical quark masses.

In principle, one could also expand the $K$-matrix, see, e.g., Refs.~\cite{Badalian:1981xj,Hyams:1973zf}.
However, such an expansion contains pole terms, which makes the fitting to data more complicated, although not impossible. In fact both parameterizations of the $K$-matrix have been recently used in the lattice study of the resonances in the coupled $\pi K-\eta K$ system \cite{Dudek:2014qha,Wilson:2014cna}.

The procedure to determine the resonance pole position consists in the following steps:
\begin{itemize}
\item[a)] The $K$-matrix is numerically extracted on the lattice, by applying the 
          L\"uscher approach;
\item[b)] The parameters $A,\, B,\dots$ are fitted to lattice data;
\item[c)] Eq.~(\ref{pole-position}) is solved on each unphysical Riemann sheet. The complex solution, which is numerically closest to the $\pi K,\,\eta K$ thresholds is identified with the $K^*$ resonance pole.
\end{itemize}
Next, we assume that the $K^*$ resonance is located on the sheet II. Other cases can be studied along the same lines.

\subsection{Pole extraction of the form factors}
We proceed with the evaluation of two- and three-point functions in the infinite volume. Afterwards, the result will be analytically continued to the resonance pole. The two-point function in Minkowski space is given by
\begin{equation}\label{2point-pole}
i\,\langle0|T [O(x) O^{\dagger}(y)]|0\rangle=\int\frac{d^4P}{(2\pi)^4}\,e^{-iP(x-y)}D(P^2),
\end{equation}
where the function $D(P^2)$ reads
\begin{equation}\label{2-point-infinite}
D(P^2)=X^T[G_{II}(s)+G_{II}(s) T_{II}(s) G_{II}(s)]X.
\end{equation}
Here, $P^2=s$ and the loop function $G_{II}(s)$ is chosen as
\begin{equation}
G_{II}(s)=
\begin{pmatrix}
-\frac{ip_1}{8\pi\sqrt{s}} & 0\\
0 & \frac{ip_2}{8\pi\sqrt{s}}
\end{pmatrix}.
\end{equation}
The form of the $G_{II}$ guaranties that the scattering amplitude $T$, which is obtained from the Lippmann-Schwinger equation,
\begin{equation}
T_{II}=(V^{-1}-G_{II})^{-1},
\end{equation}
has poles on the sheet II. The simplest way to determine the $T_{II}$-matrix is to make the replacements $\tau_1\rightarrow\,-i$, $\tau_2\rightarrow\,+i$ in Eqs.~(\ref{2T_L},\ref{L-equation}). 
We get
\begin{equation}\label{T-matrix_L}
T_{II}=\frac{8\pi\sqrt{s}}{h(E)}
\begin{pmatrix}
\frac{1}{p_1}[t_1(1-it_2)+\se^2 t] &-\frac{1}{\sqrt{p_1p_2}}\ce\se  t\\ 

-\frac{1}{\sqrt{p_1p_2}}\ce\se  t& \frac{1}{p_2}[t_2(1+it_1)-\se^2 t]
\end{pmatrix},
\end{equation}
where the quantity $h(E)$ is given by 
\begin{equation}
h(E)\equiv(t_1-i)(t_2+i)+2i\se^2(t_2-t_1).
\end{equation}
The resonance pole position $E=E_R\equiv\sqrt{s_R}$ is obtained from the equation
\begin{equation}\label{seculareq}
 h(E_R)=0.
\end{equation}
Inverting the integral Eq.~(\ref{2point-pole}) and performing the integration over all variables, we get
\begin{equation}\label{2-matrix element}
D(P^2)=\frac{Z_R}{s_R-P^2},\quad
\end{equation}
where $Z_R$ is the (complex) wave-function renormalization constant of the resonance. From Eq.~(\ref{2-matrix element}) it follows that
\begin{equation}
Z_R=\lim_{P^2\rightarrow s_R}(s_R-P^2)\,D(P^2).
\end{equation}

On the other hand, the $T(s)$-matrix on the second Riemann sheet has a pole at 
$P^2=s_R$. In the vicinity of the pole, one has 
\begin{equation}\label{T_L-pole}
T_{II}^{\alpha\beta}(s)=\frac{h_\alpha h_\beta}{s_R-P^2}+\cdots.
\end{equation}
Here, the quantities $h_1, h_2$ are given by

\begin{eqnarray}
h_1^2=-\frac{8\pi\sqrt{s}}{p_1}\frac{2E(t_2+i-\se^2 t)}{h'(E)}\bigg|_{E=E_R},\quad
h_2^2=-\frac{8\pi\sqrt{s}}{p_2}\frac{2E(t_1-i+\se^2 t)}{h'(E)}\bigg|_{E=E_R},
\end{eqnarray}
where $h'(E)\equiv dh(E)/dE$. Consequently, we obtain the renormalization constant $Z_R$:
\begin{equation}
Z_R=-\frac{1}{64\pi^2E_R^2}\biggl [\sum_{\alpha=1}^{2} (-1)^\alpha\,X_\alpha p_\alpha(E_R)h_\alpha(E_R)\biggr]^2.
\end{equation}

The calculation of the three-point function in the infinite volume proceeds in a similar manner. One gets
\begin{equation}
\label{3point-pole}
i\,\langle 0|T[O(x)J^M(0)] |B(p)\rangle=\int\frac{d^4P}{(2\pi)^4}\,e^{-iPx}\Gamma^M(P,p),
\end{equation} 
where the quantity $\Gamma^M(P,p)$  in the frame $P^\mu=(P_0,{\bf 0})$, $p^\mu=(\sqrt{m_B^2+{\bf q}^2},\,{\bf q})$ reads
\begin{equation}\label{3-point-pole}
\Gamma^M(P,p)=X^T[G_{II}(s)+G_{II}(s) T_{II}(s) G_{II}(s)]\bar F^M(P_0,|{\bf q}|).
\end{equation} 
Further, recall that the irreducible amplitudes $\bar F^M_\alpha(P_0,|{\bf q}|),\,\alpha=1,2$, are analytic functions in the complex energy plane. Then,
following Refs.~\cite{Mandelstam,Huang-Weldon}, in which the case of matrix elements between the bound states has been first studied, we {\it define} the current matrix elements at the resonance pole as
\begin{equation}\label{FF-pole definition}
F_R^M=\lim_{P^2\rightarrow s_R}Z_R^{-1/2}(s_R-P^2)\,\Gamma(P,p).
\end{equation}
Using Eqs.~(\ref{T_L-pole},\ref{3-point-pole},\ref{FF-pole definition}), we arrive at the final result:
\begin{equation}\label{ff-pole}
F^M_R(E_R,|{\bf q}|)=-\frac{i}{8\pi E}\big(p_1h_1{\bar F}^M_1-p_2h_2{\bar F}^M_2\big)\bigg|_{E=E_R}.
\end{equation}
Note that one still has an overall sign ambiguity in this formula. The corresponding form factors can be read off from Eq.~(\ref{7ffs}), in which the kinematic factors are low-energy polynomials.

In order to reproduce the one-channel result of Ref.~\cite{Agadjanov:2014kha}, the mixing between the channels should be neglected. Then, $h(E)$ takes the form
\begin{equation}
h(E)=(t_1-i)(t_2+i).
\end{equation}
So, one has at the pole position either $t_1(E_R)=+i$ or $t_2(E_R)=-i$. Consider, for instance, the first alternative $t_1(E_R)-i=0$. The derivative $h'(E)$ at $E=E_R$ reads
\begin{equation}
h'(E_R)=(t_2(E_R)+i)t_1'(E_R),
\end{equation}
so that the quantities $h_1,\,h_2$ are given by
\begin{equation}
h_1^2=-\frac{16\pi E^2}{p_1t_1'(E)}\bigg|_{E=E_R},\quad h_2^2=0.
\end{equation}
Consequently, from Eq. (\ref{ff-pole}) we obtain
\begin{equation}\label{ff-1channel}
F^M_R(E_R,|{\bf q}|)=\sqrt{\frac{p_1}{4\pi \,t_1'(E)}}{\bar F}_1^M(E,|{\bf q}|)\bigg|_{E=E_R}.
\end{equation}
A similar formula holds for the $K\eta$ channel.

\subsection{Photon virtuality}

The analytic continuation to the resonance pole yields the quantity
$F^M_R(E_R,|{\bf q}|)$. Below, we would like to briefly discuss a few conceptual
issues, related to the interpretation of this quantity. Namely, we wish
to know:
\begin{itemize}
\item
What is the photon virtuality $q^2$  for the resonance form factor, extracted
at the pole?
\item
How should one compare with the experimental results?
\end{itemize} 
In the literature, different statements have been made on this issue so far. 
We think that a clarification is needed at this point.

According to the procedure, which is proposed in the present paper (see also Ref.~\cite{Agadjanov:2014kha}), the finite-volume
matrix element is measured at different two-particle energies $E_n(L)$ and
a fixed value of $|{\bf q}|$. After that, an analytic continuation is performed
to the complex resonance pole, keeping $|{\bf q}|$ fixed. Further, the
photon virtuality becomes complex at the pole
\eq\label{eq:virt}
q^2=\bigg(E_R-\sqrt{m_B^2+{\bf q}^2}\bigg)^2-{\bf q}^2\, .
\en

\begin{figure}[placement]
\begin{center}
\includegraphics[width=7.cm]{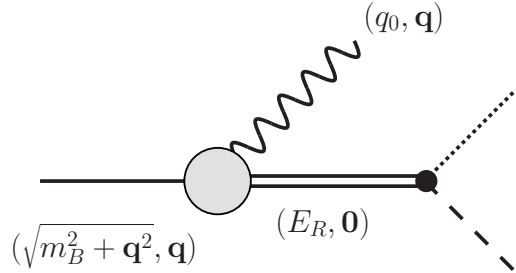}
\end{center}
\caption{\small The factorization of the amplitudes at the resonance
pole (see Fig.~2 for notations). The photon virtuality, given by Eq.~(\ref{eq:virt}), is complex.
}
\label{fig:factorization}
\end{figure}

On the other hand, in Refs.~\cite{Briceno:2015dca,Briceno:2016kkp}, where the  $\rho\to\pi\gamma^*$ transition form factor is considered, the authors simultaneously 
parameterize the energy- and 
$q^2$-dependence of the measured matrix element by some phenomenological fit function
and perform the analytic continuation to the complex value of energy at 
a fixed $q^2$. The quantity $q^2$ is taken real at the pole.

Having two different procedures for the determination of the matrix element at the pole, it seems that the result is not unique. In order to show that the form factor
can be uniquely defined, we note that the residue of the full amplitude at the pole should
factorize in the product of the resonance form factor
and the vertex, describing the transition of a resonance
into the final state, see Fig.~\ref{fig:factorization}. The background becomes irrelevant, which leads to the determination of the form factor at the pole in a process-independent manner. From this figure
it is clear that the photon virtuality, defined through the use of the 
4-momentum conservation, coincides with the one given in Eq.~(\ref{eq:virt})
and thus must be complex. One could of course consider the electroproduction 
amplitude at a different (even at a real) photon virtuality as well. However, 
in this case, the background does not vanish completely, so the continuation
to the pole does not make sense, since the result is process-dependent anyway.
It should be stressed that this argument equally holds both in the analysis
of the data from the electroproduction experiments as well as for the results of lattice QCD simulations.

Another argument addresses the analytic properties of the amplitudes which are
extrapolated into the complex plane. We have shown
that the irreducible amplitudes are low-energy polynomials 
in the vicinity of a resonance in the CM energy $E$, if 
the photon 3-momentum $|{\bf q}|$ is fixed (see Ref.~\cite{Agadjanov:2014kha}). This fact implies that the analytic 
continuation to the complex energies is robust. To the best of our knowledge,
no such statement exists 
in case of the function of two independent variables $E$, $q^2$ that might
render the analytic continuation unstable. It remains to be seen, whether
the information about the analytic properties of the form factors in the 
variable $q^2$ can be reasonably included. This could greatly constrain the 
fit and would be very useful in the analysis of the presently available data,
which correspond to different values of  $q^2$ (see, e.g., 
Refs.~\cite{Briceno:2015dca,Briceno:2016kkp}).

\section{Infinitely narrow width}

In this section, for illustrative purposes,  we consider the case of a resonance 
with an infinitely narrow width  having  in mind the hypothetical case of a  
$K^*$ pole located above the $\eta K$ threshold with a very small width.
The arguments follow the path of Ref.~\cite{Agadjanov:2014kha}, where the
same problem has been considered in case of the elastic scattering (see also Ref.~\cite{Briceno:2015csa}). 
It has been shown there that, in the limit of the infinitely narrow width,
the matrix element, measured on the lattice, coincides with the infinite-volume
resonance form factor up to a constant, which takes into account the 
difference between the normalization of the one- and two-particle states in a finite volume. However, the multi-channel
case is more subtle, since different two-particle states occur, 
and the relation between the infinite- and finite-volume 
matrix elements becomes
obscure. Still, as we will see, the final result has exactly the same form as in the one-channel problem\footnote{Inadvertently, in Ref. \cite{Agadjanov:2014kha}, the factor ${\cal V}^{-1/2}$ was missing on the right-hand side of the counterpart of Eq. (\ref{3point-integral}).}.    

We start with the two-body
potential from Eq.~(\ref{potential}), which can be written in the following form
\eq
V=8\pi\sqrt{s}P^{-1/2}O\tilde VO^TP^{-1/2}\, ,
\en
where
\eq
P=\mbox{diag}\,(p_1,p_2)\, ,\quad\quad
\tilde V=\mbox{diag}\,(t_1,t_2)\, ,\quad\quad
O=\begin{pmatrix}\ce & -\se \cr \se & \ce\end{pmatrix}\, .
\en
Suppose that the resonance behavior near the (real) energy $E=E_0$ emerges in the 
quantity $t_1=\tan\delta_1$, whereas the quantity $t_2$ stays regular in this
energy interval. Then, in the vicinity of $E=E_0$, one can write
\eq
\delta_1(E)=\delta_R(E)+\phi(E)\, ,\quad\quad
\tan\delta_R(E)=\frac{\Gamma_0/2}{E_0-E}\, ,
\en
and assume that a (small) background phase $\phi(E)$ stays regular.
Further, one may straightforwardly ensure that
\eq\label{eq:t1}
\cot\delta_1(E)=\frac{ E_{BW}-E}{\Gamma/2}+\cdots\, ,
\en
where
\eq
E_{BW}=E_0-\frac{\Gamma_0}{2}\,\tan\phi(E_{BW})\, ,\quad\quad
\frac{\Gamma}{2}=\frac{\Gamma_0}{2}\,(1+\tan^2\phi(E_{BW}))\, .
\en
This shows that, in the vicinity of a narrow resonance, one can always
get rid of the background phase by a redefinition of the resonance parameters. We note that the second background phase still remains.
The quantities $ E_{BW}$ and $\Gamma$ are the Breit-Wigner mass and 
width of the (narrow) resonance.

In the vicinity of a narrow resonance, the scattering amplitude Eq.~(\ref{T-matrix}), which can be represented on the first Riemann sheet as
\eq
T=8\pi\sqrt{s}P^{-1/2}O\tilde TO^TP^{-1/2},\quad \tilde T=\mbox{diag}\,(e^{i\delta_1}\sin\delta_1,e^{i\delta_2}\sin\delta_2),
\en
becomes
\eq\label{eq:T}
T_{\alpha\beta}=\frac{b_\alpha b_\beta}{s_{BW}-s-i\sqrt{s_{BW}}\,\Gamma}+\mbox{regular terms at $E\to E_{BW}$}\,,
\en
where $s_{BW}=E_{BW}^2$. Here, the quantities $b_1,\,b_2$ are given by 
\begin{equation}
b_1=\sqrt{\frac{8\pi s_{BW}\Gamma}{p_1}}\,\ce\, ,\quad 
b_2=\sqrt{\frac{8\pi s_{BW}\Gamma}{p_2}}\,\se\, ,
\end{equation}
and the regular terms emerge from the contribution of $t_2$.

In order to find a complex pole on the second Riemann sheet, one has to solve the 
secular equation, Eq.~(\ref{seculareq}), $h(E_R)=0$. Recalling that $t_1,t_2$ are single-valued 
functions and using the explicit representation of $t_1$ from 
Eq.~(\ref{eq:t1}), at $E=E_R$ we get
\eq
t_1(E_R)=\frac{\Gamma/2}{E_{BW}-E_R}
=\frac{i(t_2+i)-2i\se^2t_2}{t_2+i-2i\se^2}\bigg|_{E=E_R}.
\en

\subsection{Real axis}
\begin{sloppypar}
On the real energy axis, one can introduce the infinite-volume quantities (``form factors''), which parameterize the imaginary parts of the decays amplitudes ${\cal A}^M_1,\,{\cal A}^M_2$ in the vicinity of the Breit-Wigner resonance.
We denote these volume-independent matrix elements as $F_A^M(E,|{\bf q}|)$. In analogy to the one-channel case (see, e.g. Refs.~\cite{Aznauryan,Drechsel,Briceno:2015csa}), we consider the resonance exchange mechanism at tree level, as shown in Fig. 4. Consequently, the amplitudes ${\cal A}^M_1,\,{\cal A}^M_2$  near $E=E_{BW}$ read
\end{sloppypar}
\begin{equation}
{\cal A}^M_\alpha(E,|{\bf q}|)=\frac{b_\alpha F_A^M(E_{BW},|{\bf q}|)}{E^2_{BW}-E^2-iE_{BW}\Gamma}+\cdots\,,\quad \alpha=1,2,
\end{equation}
where the ellipses stand for the terms emerging from the regular contributions
in Eq.~(\ref{eq:T}).
Setting further $E=E_{BW}$, we get the imaginary parts of the ${\cal A}^M_\alpha$ 
\eq\label{amplitudes-narrow}
{\rm Im}{\cal A}^M_1(E_{BW},|{\bf q}|)&=&\sqrt{\frac{8\pi}{p_1\Gamma}}F_A^M(E_{BW},|{\bf q}|)\ce+O(1)\, ,
\nonumber\\[2mm]
{\rm Im}{\cal A}^M_2(E_{BW},|{\bf q}|)&=&\sqrt{\frac{8\pi}{p_2\Gamma}}F_A^M(E_{BW},|{\bf q}|)\se+O(1)\, . 
\en
Note that the leading terms in this expression are of order $\Gamma^{-1/2}$, and
the sub-leading $O(1)$ terms emerge from the regular contributions.

\begin{figure}[placement]
\begin{center}
\includegraphics[scale=0.5]{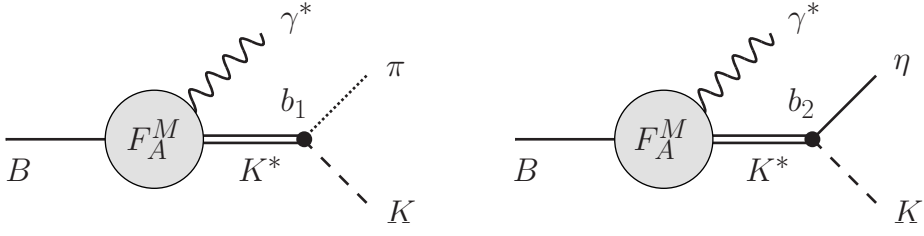}
\end{center}
\caption{\small Decay amplitudes ${\cal A}^M_\alpha,\,\alpha=1,2$, in the vicinity of the infinitely narrow $K^*$. The quantities $b_\alpha,\,\alpha=1,2$, denote the couplings of the $K^*$ to the respective channels at $E=E_{BW}$.}
\end{figure}

Further, comparing Eqs.~(\ref{Watson}) and (\ref{amplitudes-narrow}), we see that the following condition has to be satisfied at the Breit-Wigner pole $E=E_{BW}$: 
\begin{equation}
u_2^M(E_{BW},|{\bf q}|)=O(1),
\end{equation}
while for the amplitudes $u_1^M(E_{BW},|{\bf q}|)$ one has
\begin{equation}
u_1^M(E_{BW},|{\bf q}|)=\sqrt{p_1}\ce\,{\rm Im}{\cal A}^M_1(E,|{\bf q}|)\big|_{E_n\rightarrow E_{BW}}+\sqrt{p_2}\se\,{\rm Im}{\cal A}^M_2(E,|{\bf q}|)\big|_{E_n\rightarrow E_{BW}},
\end{equation}
or
\begin{equation}\label{u1-narrow}
u_1^M(E_{BW},|{\bf q}|)=\sqrt{\frac{8\pi}{\Gamma}}F_A^M(E_{BW},|{\bf q}|)+O(1)=O(\Gamma^{-1/2})\, .
\end{equation}
Consequently, in the limit $\Gamma\to 0$, the leading contribution to the ${\cal A}^M_\alpha$  comes
from $u_1^M$.

However, the amplitudes $u_\alpha^M,\,\alpha=1,2$ are not low-energy polynomials 
in the vicinity of $E=E_{BW}$. In order to establish quantities, which have such a 
property, we first note that in case of a very narrow resonance, the function 
$\cot\delta_1(E)$ is a polynomial in $E$ (see Eq.~(\ref{eq:t1})). It can be further 
assumed that the mixing parameter $s_\epsilon(E)$ and $\cot\delta_2(E)$ are also  
low-energy polynomials in the vicinity of the resonance.  Furthermore, even if the 
radius of convergence of the modified effective range expansion, Eq.~(\ref{ERE}), 
is assumed to be much larger than the width $\Gamma$, it is still limited from above 
by the distance to the nearest threshold. Since the limit $\Gamma\to 0$ is considered 
here, it is natural to assume that the mixing parameter $s_\epsilon(E)$ and 
$\cot\delta_2(E)$ are also  low-energy polynomials in the vicinity of the resonance. 
It is then straightforward to check that the functions
\eq\label{u-tilde}
\tilde u_\alpha^M=\frac{u_\alpha^M}{\sin\delta_\alpha}
\en
are low-energy polynomials. Indeed, the irreducible amplitudes $\bar F_\alpha^M,\alpha=1,2$, diverge at $E=E_{BW}$, due to the propagation of the bare $K^*$ in the $s$-channel (see Ref.\cite{Agadjanov:2014kha}). According to Eqs.~(\ref{u1u2}) and~(\ref{u-tilde}), this divergence is exactly canceled in the amplitudes $\tilde u_\alpha^M,\,\alpha=1,2$. Consequently, they can be safely expanded in the vicinity of the narrow resonance. This property, in particular, is important, if one considers an analytic continuation into the complex plane.

Rewriting the two-channel Lellouch-L\"uscher formula
in terms of $\tilde u_\alpha^M$, we get
\begin{equation}\label{2LL-alternative}
\big|F^M(E_n,|{\bf q}|)\big|= \frac{{\cal V}^{-1}}{\sqrt{8\pi E}}\,
\big(a_1  \tilde u_1^M+a_2\tilde u^M_2\big)\bigg|_{E=E_n},
\end{equation}
where the  quantities $a_1,a_2$  are given by
\eq
a_1^2&=&t_1^2\frac{t_2+\tau_2-\se^2(\tau_2-\tau_1) }{f'(E)}\bigg|_{E=E_n}\, ,
\nonumber\\[2mm]
a_2^2&=&t_2^2\frac{t_1+\tau_1+\se^2(\tau_2-\tau_1)}{f'(E)}\bigg|_{E=E_n}\ .
\en
Evaluating the quantities $a_1,a_2$ in the limit of the infinitely narrow width
is somewhat less trivial than in the one-channel case. In order to proceed 
further here, let us first recall the line of reasoning used 
in the one-channel case. In this case, the L\"uscher equation has a simple form
\eq
\delta_1+\varphi_1=n\pi\, ,\quad\quad n\in\mathbb{Z}\, ,\quad\quad
\tan\varphi_1=\tau_1\, .
\en
For sufficiently small $\Gamma$, this equation will have a solution at 
$E_n= E_{BW}+O(\Gamma)$. At this energy, the quantities
$t_1,\tau_1$ are of order $O(1)$. However, the {\em derivatives} of $t_1$ and $\tau_1$ behave differently at $E_n\to E_{BW}$.  One has $t_1'=\delta_1'/\cos^2\delta_1$
and $\tau_1'=\varphi_1'/\cos^2\varphi_1$, where $\cos^2\delta_1=\cos^2\varphi_1$,
due to the L\"uscher equation. According to Eq.~(\ref{eq:t1}), the derivative of the phase
shift $\delta_1$ diverges as $\Gamma^{-1}$ whereas $\varphi_1'$ stays finite
as $E_n\to E_{BW}$, since it is a kinematical function that does not contain any small scales of order $\Gamma$. Consequently, as $E_n\to E_{BW}$ and $\Gamma\to 0$, one may neglect $\tau_1'$ as compared to $t_1'$.

A similar argument can be carried out in the two-channel case, rewriting the L\"uscher equation in the form
\eq
\delta_1+\varphi_1=n\pi\, ,\quad\quad
\tan\varphi_1=\frac{\tau_1(t_2+\tau_2)+\se^2t_2(\tau_2-\tau_1)}
{t_2+\tau_2-\se^2(\tau_2-\tau_1)}\, .
\en
The function $\varphi_1$ is not purely kinematical as it contains $t_2$.
However, it still does not contain small scales of order $\Gamma$. Consequently,
the derivatives of $\varphi_1$ are finite and the quantities 
$\tau_1',\tau_2'$ are of order $O(1)$, while $t_1'=O(\Gamma^{-1})$.

Next, retaining only the most divergent terms in $f'(E_n)$ at 
$E_n\to E_{BW}$, one gets
\begin{equation}
f'(E_n\rightarrow E_{BW})=t_1'\big(t_2+\tau_2-\se^2(\tau_2-\tau_1)\big)\big|_{E_n\rightarrow E_{BW}}+\cdots\, .
\end{equation}
Consequently, the quantities $a_1^2,a_2^2$ take the values
\eq
a_1^2&=&\frac{t_1^2}{t_1'}\bigg|_{E_n\to E_{BW}}=\frac{\Gamma}{2}+O(\Gamma^2) ,
\nonumber\\[2mm]
a_2^2&=&\frac{t_2^2}{t_1'}\,\frac{t_1+\tau_1+\se^2(\tau_2-\tau_1)}{t_2+\tau_2-\se^2(\tau_2-\tau_1)}\bigg|_{E_n\to E_{BW}}=O(\Gamma).
\en

 Hence, it follows that the leading contribution to the matrix element
$F^M(E_n,|{\bf q}|)$ in the limit $\Gamma\to 0$ comes only from the term, proportional to $\tilde u^M_1$, whereas the second term is sub-leading. As a result, we obtain
\begin{equation}
\big|F^M(E_n,|{\bf q}|)\big|=\frac{{\cal V}^{-1}}{\sqrt{2E_{n}}}\,\big|F_A^M(E_{n},|{\bf q}|)\big|+O(\Gamma^{1/2}),\quad E_n = E_{BW}+O(\Gamma).
\end{equation}
As seen, the Lellouch-L\"uscher formula has a fairly simple form in the 
vicinity of the Breit-Wigner resonance: the 
infinite-volume quantities 
$F_A^M(E_{BW},|{\bf q}|)$ are equal to the current matrix elements 
$F^M(E_{BW},|{\bf q}|)$, measured on the lattice , up to a normalization 
factor (note that, in Ref.~\cite{Agadjanov:2014kha}, a different 
normalization 
of the states has been used).
The form factors can be found from Eq.~(\ref{7ffs}).

\subsection{Complex plane}
The values of the form factors at the resonance pole in the infinitely narrow width limit can be determined along the same lines, as discussed above. 
We again express the final result Eq.~(\ref{ff-pole}) through the 
amplitudes $\tilde u_1^M,\,\tilde u_2^M$ to get
\begin{equation}\label{ff-pole-alternative}
F^M_R(E_R,|{\bf q}|)= \frac{1}{\sqrt{4\pi}}\,\big(r_1 \tilde u_1^M+r_2\tilde u^M_2\big)\big|_{E=E_R}.
\end{equation}
Here, the  quantities $r_1,r_2$  read
\eq\label{f1f2}
r_1^2&=&t_1^2\frac{t_2+i-2i\se^2 }{h'(E)}\bigg|_{E=E_R}\, ,
\nonumber\\[2mm]
r_2^2&=&t_2^2\frac{t_1-i+2i\se^2}{h'(E)}\bigg|_{E=E_R}\, .
\en
Since the functions $\tilde u_\alpha^M$ are low-energy polynomials in 
the vicinity of the Breit-Wigner pole, one can analytically 
continue them from the real axis to the pole.  Consequently, in the limit $\Gamma\to 0$, their values at the pole and at the real axis are equal, up to the terms of order $O(\Gamma)$. We note that
this procedure cannot be applied to the $u^M_\alpha$. Calculating the quantities $r_1,r_2$ at $E_R\to E_{BW}$, we get
\eq
r_1^2&=&\frac{t_1^2}{t_1'}\bigg|_{E_R\to E_{BW}}=\frac{\Gamma}{2}+O(\Gamma^2)\, ,
\nonumber\\[2mm]
r_2^2&=&\frac{t_1^2}{t_1'}\,\frac{t_2+i-2i\se^2 }{t_1-i+2i\se^2}\bigg|_{E_R\to E_{BW}}=O(\Gamma).
\en
As on the real axis, the leading contribution to the $F^M_R$ is dominated by the 
$\tilde u_1^M$ term in Eq. (\ref{ff-pole-alternative}). The final expression takes the form
\begin{equation}
F^M_R(E_R,|{\bf q}|)\big|_{\Gamma\to 0}=F_A^M(E_{BW},|{\bf q}|)+O(\Gamma^{1/2}).
\end{equation}
As expected, for infinitely narrow resonance, the form factors $F_A^M(E,|{\bf q}|)$ and $F_R^M(E,|{\bf q}|)$, defined on the real energy axis and complex plane, respectively, coincide.

\section{Conclusions}
In this work, we have studied the extraction of the  
$B\to K^*$ transition form factors on the lattice. We have
 taken into account, in particular, the possible admixture of the 
$\eta K$ to $\pi K$ final states. To this end, we have applied the 
non-relativistic effective field theory in a finite volume 
and reproduced the two-channel analogue of the Lellouch-L\"uscher formula, 
which allows to extract the $B\to K^*l^+l^-$ decay amplitude in the 
low-recoil region. 

Since the $K^*$ is a resonance, the corresponding current matrix elements 
are properly defined and free of process-dependent ambiguities only if 
the analytic continuation in the complex energy plane to the resonance 
pole position is performed. Consequently, we have set up a framework for 
the determination of the form factors at the $K^*$ pole. This is a 
generalization of 
the one-channel formula, which has been derived in Ref.~\cite{Agadjanov:2014kha}.
In addition, we have discussed in detail the consistent determination of the photon
virtuality at the resonance pole.  

Finally, we have considered the limit of an infinitely small width in our 
results. The equations in the multi-channel case are more involved and this 
limit cannot be performed in a straightforward manner. Nevertheless, we have demonstrated that, even in the multi-channel case, the current matrix element measured on the lattice is equal to the one in the infinite volume, up to a normalization factor that does not depend on the dynamics. This result represents a useful check of our framework.

\bigskip\bigskip\bigskip

\noindent{\it Acknowledgments:} We thank R. Brice\~no and M. Mai for useful discussions. 
We acknowledge the support by the DFG (CRC 16,
``Subnuclear Structure of Matter'' and CRC 110 ``Symmetries and the Emergence of Structure in QCD'') and by the Bonn-Cologne Graduate School of Physics and
Astronomy. This research is supported in part by Volkswagenstiftung
under contract no. 86260, and by the Chinese Academy of Sciences (CAS) President’s
International Fellowship Initiative (PIFI) (Grant No. 2015VMA076).

\bigskip\bigskip\bigskip

\appendix

\numberwithin{equation}{section}
\setcounter{equation}{0}
\setcounter{figure}{0}
\setcounter{table}{0}

\section{The {\boldmath$B\to K^*$} form factors in rest frame of the {\boldmath$B$} meson }
Since the $\pi K-\eta K$ system is in the P-wave, it is preferable to extract the finite-volume energy spectrum in the reference frame, in which the $K^*$ has non-zero 3-momentum. Consequently, as an alternative to the case considered in the main text, we also consider the following kinematics:
\begin{equation}\label{kinematics}
{\bf p}=0,\quad {\bf k}=-{\bf q}=\frac{2\pi}{L}{\bf d},\quad {\bf d}=(0,0,n)~.
\end{equation}
The current matrix elements of Eq.~(\ref{7ffs}) in this moving frame take the form
\begin{eqnarray}\nonumber\label{7ffs-moving}
\langle V(k,+) | J^{(+)}| B\rangle
&=& -\frac{2 i m_B|{\bf q}|V(q^2)}{m_B + m_V},\\[2mm]
\langle V(k,0) | i(m_B-E_V) J_A + |{\bf q}| J^{(0)}_A | B\rangle
&=& - 2im_B|{\bf q}| A_0(q^2),\nonumber\\[2mm]
\langle V(k,+)|J^{(+)}_A| B\rangle
&=& -i(m_B+m_V) A_1(q^2),\nonumber\\[2mm]
\langle V(k,0) |   i(m_B-E_V)J^{(0)}_A-|{\bf q}| J_A | B\rangle
&=&  8m_Bm_V A_{12}(q^2),\nonumber\\[2mm]
\langle V(k,+) | i(m_B-E_V)I^{(+)} + |{\bf q}|I^{(+)}_0 | B\rangle
&=&  2im_B|{\bf q}| T_1(q^2),\nonumber\\[2mm]
\langle V(k,+) | i(m_B-E_V)I^{(+)}_{A} + |{\bf q}|I^{(+)}_{0A} | B\rangle
&=&  -i(m_B^2-m_V^2)T_2(q^2),\nonumber\\[2mm]
\langle V(k,0) | 
I^{(0)}_A| B\rangle
&=& -\frac{4m_Bm_V}{m_B+m_V} T_{23}(q^2),
\end{eqnarray}
where $E_V=\sqrt{m^2_V+{\bf q}^2}$ is energy of the $K^*$ meson, which is treated as a stable particle. As seen from Eq.~(\ref{kinematics}) (see, e.g., also Ref.~\cite{Gockeler:2012yj}), the matrix elements should be measured in the irreps $\mathbb{E}$ and $\mathbb{A}_1$ of the little group $C_{4v}$. However, because $K^*$ is not at rest now, the S- and P-waves mix in the irrep $\mathbb{A}_1$. Consequently, only the form factors $V$,\,$A_1$,\,$T_1$, and $T_2$ could be extracted by applying formulas that are similar to the ones given in the previous sections. For other form factors  $A_0$,\,$A_{12}$,\,$T_{23}$ one should either assume that the mixing is small, or use more general equations, derived in Ref.~\cite{Briceno:2014uqa}, which include it.

Further, one applies the following operators to create the states $\langle V(k,\pm) |,\,\langle V(k,0) |$ from the vacuum:
\begin{eqnarray}
{\cal O}_{\mathbb{E}}^{(\pm)}({\bf k},t)=\frac{1}{\sqrt{2}}\sum_{\bf x}e^{i{\bf k}{\bf x}}\big(O_1({\bf x},t)\mp iO_2({\bf x},t)\big),\quad {\cal O}_{\mathbb{A}_1}^{(0)}({\bf k},t)=\sum_{\bf x}e^{i{\bf k}{\bf x}}O_3({\bf x},t).
\end{eqnarray}

When the $K^*$ becomes unstable, the mass $m_V$ should be replaced by the CM energy value $E_n^*$ of the two-particle state in a finite volume. Then, the matrix elements Eq.~(\ref{7ffs-moving}) are functions of $E_n^*$ and $|{\bf q}|$. Analogously, in order keep $|{\bf q}|$ fixed at different values of energy $E_n^*$, one could resort to asymmetric volumes of type $L\times L\times L'$ or twist the $s$-quark.

\end{document}